\documentclass[11pt]{article}

\usepackage{color}
\usepackage{graphicx}
\usepackage{amssymb}
\usepackage{amsmath}
\usepackage{colortbl}
\usepackage{color}
\usepackage{bbding}
\usepackage{url}
\usepackage{multirow}
\usepackage{times}
\usepackage{fullpage}
\usepackage{indentfirst}
\usepackage[bf]{caption} 
\usepackage{subfigure}
\usepackage[hang,flushmargin,stable]{footmisc}

\usepackage{flushend} 
\usepackage[top=0.9in, bottom=0.9in, left=0.9in, right=0.9in]{geometry}

\makeatletter
\def\url@leostyle{%
  \@ifundefined{selectfont}{\def\UrlFont{\sf}}{\def\UrlFont{\small\ttfamily}}}
\makeatother	
\newcommand{\remove}[1]{}
\def\naive{na\"{i}ve}
\usepackage{ifthen}
\newcommand{\longver}[1]  {\ifthenelse{\equal{\shortversion}{false}}{{#1}}{}}
\newcommand{\descr}[1]{\vspace{0.25cm}\noindent \textbf{#1}}

\usepackage{footmisc}

\newcommand{\modified}{}

\title{\bf EphPub: Toward Robust Ephemeral Publishing\thanks{An earlier version of this paper appeared in the Proceedings of
IEEE ICNP 2011.}\vspace{0.4cm}}

\author{Claude Castelluccia$^{1}$, Emiliano De Cristofaro$^{2}$, Aurelien Francillon$^3$, Mohamed-Ali Kafaar$^1$
\vspace{0.2cm}\\
{\normalsize $^1$ INRIA Rhone Alpes, France}\\
{\normalsize $^2$ University of California, Irvine \& PARC}\\
{\normalsize $^3$  ETH Zurich, Switzerland \& Institute Eurecom, France}}

\date{}
\begin{document}
\maketitle

\begin{abstract}
  The increasing amount of personal and sensitive information disseminated over the Internet
  prompts commensurately growing privacy concerns. Digital data often lingers indefinitely and
  users lose its control.  This motivates the desire to restrict 
  content availability to an {\em expiration} time set by the data owner.
  This paper presents and formalizes the notion of {\em Ephemeral Publishing} (EphPub),
  to prevent the access to expired content.  
  We propose an efficient and robust protocol that builds on the Domain Name System (DNS)
  and its caching mechanism. With EphPub, sensitive content is published {\em encrypted} 
  and the key material is distributed, in a steganographic manner, to randomly selected and independent resolvers.
  The availability of content is then limited by the evanescence of DNS cache entries.
  The EphPub protocol is transparent to existing applications, and does not rely on trusted hardware,
  centralized servers, or user proactive actions.
  We analyze its robustness and show that it incurs a negligible overhead on the DNS infrastructure.
  We also perform a large-scale study of the caching behavior of 900K open DNS resolvers.
  Finally, we propose Firefox and Thunderbird extensions that provide ephemeral publishing capabilities, as well as a command-line tool to
  create ephemeral files.
\end{abstract}

\section{Introduction} \label{Introduction}
As the amount of private information disseminated over the Internet increases, so do related privacy concerns. 
Private data is increasingly often cached, stored, or archived,  {\em in the cloud} and,
in numerous occasions, its ownership is lost.
Users are often unable to successfully delete published content:
for instance, social network providers maintain daily backup tapes and do not physically 
erase ``deleted'' content from all backups~\cite{facebook}, ascribing to 
the expensiveness of this operation. As a result, everlasting information may  endanger user privacy
and become a prey of subpoena, government surveillance, or data leaks.
For instance, a college student may upload (e.g., on a social network) pictures 
where she/he and her/his friends are evidently drunk; even after years, pictures may regrettably re-surface, e.g., during job hunting.
Similarly, a blogger may post some controversial messages, which can later compromise an election campaign or even personal safety. 
In essence, if private data is perpetually available, then the threat for user privacy becomes permanent.

      One natural step toward enhancing privacy is ensuring data
      confidentiality, i.e., by encrypting sensitive content. However,
      besides the well-known key-distribution problem, this approach faces
      some challenges: (1) privacy is lost if cleartext copies of
      the sensitive content or the decryption keys are leaked, and (2) access
      cannot be revoked. In particular key holders  who
      had little
      interest in the data at the time of distribution can later decide to
      use the key to access the data at any future time. Consider, for
      example, a friend that has become an enemy, an employer that wants to
      dig through an employee's past data, or subpoenas for old
      records. Unfortunately, not much can be done with respect to the
      first problem where content or keys that are leaked or
      redistributed. In this paper we address the later problem; namely
      the problem of revoking encryption keys from key holders who try to
      access the data after some future expiration time set by the user.

Our approach, called {\em Ephemeral Publishing (EphPub)}, not only encrypts
      content, but also uses a key escrow mechanism that ensures a key’s
      automatic erasure, after an expiration time set by the user.  Our
      approach is similar to the concept explored by Vanish~\cite{Vanish}, where
      key shares are distributed on a Distributed Hash Table (DHT).  Since
      peers store data only for a limited period of time, so called churn,
      key shares are eventually erased. However, DHTs are
      vulnerable to Sybil attacks~\cite{Sybil}. It has been shown by Wolchok
      et. al.~\cite{Defeat-Vanish} that, at a reasonable cost, an attacker can collect the
      keys distributed over the DHT.  As we will show, EphPub uses a novel
      key distribution mechanism based on the DNS system that avoids such
      attacks and also improves the usability and availability of the
      system.

\smallskip\noindent The notion of Ephemeral Publishing is useful, for instance, in the following scenarios:

\begin{itemize}
\item {\bf\em Online Social Networks (OSNs):} Content published on
  online social networks---such as pictures or wall posts---has often
  time-limited value to the users. As reported by~\cite{zepho}, users
  would often like to delete their content after a small period of
  time (e.g., a few hours or days).  
  \modified{However, deletion is not always effective, as data is often
    cached without users' knowledge by the service provider~\cite{zombie}.  In
    contrast, EphPub lets a user publish encrypted content and ensures
    that intended recipients (e.g., OSN friends) can access decryption
    keys -- hence, content -- only within a specific period of time,
    assuming the OSN friend has not cached the decryption key for
    future use.  While EphPub cannot prevent OSN friends from caching
    content or decryption keys for future access to the data, it can
    prevent friends from viewing data in the future if they did not
    cache the key or the content (e.g., they had no interest in the
    data at the time and did not access the content or key)\footnote{ Our implementation of EphPub 
      never caches unencrypted keys or content, which means users cannot
      access the content after the key expires.}.
  }

\item   {\bf\em Outsourced Data:} Web users increasingly outsource
    personal and confidential data to the cloud, e.g., using backup
    services, such as Dropbox.  
    \modified{ Such private data must be stored in an
    encrypted form to prevent access by the service provider (e.g.,
    for data mining purposes or in response to a subpoena).  However,
    encryption would not be effective unless all the users of the data
    proactively delete the decryption keys since the keys may be
    subject to subpoena.  Note that in this case we assume there is no
    risk of leaks or redistribution because all users want to keep the
    data private.  However, it requires all users to remember to
    delete the decryption keys they have been given.  
    EphPub addresses this issue by distributing information that
    allows users to retrieve the key online. This information does
    not need to be deleted because it becomes useless after a fixed
    amount of time.}
\end{itemize}

A \naive\ solution for the EphPub problem would  be to publish encrypted content and
store the corresponding encryption keys on a server that she owns 
so that the keys can be deleted when the content's lifetime expires.
However, this solution requires keys to be constantly accessible 
(thus, introducing a single point of failure), and raises additional privacy concerns as the publisher can always trace
users accessing her content.

\descr{Technical Roadmap.} In this paper we present Ephemeral Publishing (EphPub), a protocol aimed at guaranteeing {\em retroactive privacy}~\cite{Vanish},
i.e., allowing users to set an expiration time on their information and
preventing an adversary from retrieving messages {\em after} expiration. 
EphPub leverages a fully-distributed and ubiquitous Internet service: the Domain Name System (DNS).
We exploit its caching mechanism, as DNS resolvers cache the response to a recursive DNS query for potential further requests. These cache entries are kept for a fixed period of time, i.e., the record's TTL (Time To Live). Once the TTL has expired, the resolver erases the record from its cache. 
We implement encryption keys as {\em ephemeral keys}. 
Each key is composed of {\em ephemeral bits}, i.e., it is divided into single bits that independently become inaccessible after a period of time.

\descr{Ephemeral Bits.} 
To set an ephemeral bit to $1$, a \emph{recursive} DNS query for an existing domain name is performed to a DNS resolver.
This resolver caches the response for as long as the domain's TTL permits. 
Whereas, to set an ephemeral bit to $0$, no DNS request is performed. An ephemeral bit is thereafter coded by the domain name and the resolver address. 
Before the TTL expires, one can retrieve the ephemeral bit with a \emph{non-recursive} DNS query:  if the resolver answers the query (i.e., an entry in its
cache exists) then the bit $1$ is read, otherwise if no response is returned the bit $0$ is read.
We use domain names selected at random, so that they have a negligible
probability to be queried by entities external to the system. To deal with events that may generate faulty bits (DNS churn,
connectivity problems, caching implementations, policies issues,
failures, etc.), we use a Reed-Solomon error correction code~\cite{reed-solomon}.



\descr{Contributions.} 
This paper formalizes the notion of {\em Ephemeral Publishing} (EphPub), by defining requirements, adversarial capabilities,
and privacy goals. 
Then, it presents a practical and robust protocol that uses the DNS and its caching mechanism. 
Our solution has several strong points compared to related work (e.g., Vanish~\cite{Vanish}): 
besides being immune to Sybil attacks, (1) it allows users to set data expiration times with finer granularity (in Vanish it is bounded to DHT churn, typically 8 hours); (2) it does not require any additional infrastructure (e.g., a DHT or trusted servers); (3) it is lightweight,
and does not force users to install extra software (e.g., DHT client), thus, it can be deployed on smartphones.
Further, we propose two prototype implementations: a command-line tool to create files that can be decrypted only before expiration time, 
and a stand-alone Firefox and Thunderbird extension that makes any web text ephemeral. 
Source code is available at \url{http://code.google.com/p/ephpub/}.
Finally, in the process of designing our EphPub solution, we analyze the behavior of DNS cache resolvers:
our measurements show that, from a dataset of 225K resolvers, 10\% behave as recommended by the IETF. 
We stress that the goal of EphPub is to protect information privacy against
third-party service providers, government surveillance, and future subpoena.
EphPub is {\em not} designed to safeguard published content from targeted attacks, e.g., within content's lifetime. 
It does not provide a DRM-like mechanism that prohibits recipients from copying or republishing content, {\em before}
expiration time. In other words, aiming at preventing data access {\em after} expiration time,
EphPub proposes a new automated technique (i.e., without user proactive intervention) geared for time-bounded content.

\descr{Paper Organization.} The rest of the paper is organized
  as follows.  In Section~\ref{sec:req}, we define system model, requirements, and privacy properties.
  Then, we describe the EphPub protocol in Section~\ref{EphPub_Protocol} and
  analyze its security in Section~\ref{security_analysis}. 
  Next, Section~\ref{software} presents the details of our
  prototype implementations, Section~\ref{experiments}
  presents our extensive experimental analysis, and 
  Section~\ref{related} overviews relevant related work.
  Finally, Section~\ref{conclusion} concludes the paper and discusses future work. 
  
\section{Preliminaries}
\label{sec:req}
In this section, we model system assumptions and requirements. 
Finally, we present adversarial models and our formal definition of retroactive privacy.


\descr{System Model.} An Ephemeral Publishing (EphPub) system involves a sender, $S$, and one or more receivers.
For ease of discussion, and without loss of generality, in the rest of the paper, we assume a single receiver $R$.
$S$ sends to $R$ {\em time-bounded} messages $M$. A message $M$ is {\em time-bounded} if it can only be read for a given
period of time specified by $S$, that we denote with $T_v$. 
We denote with $t$ the time at which the validity of the message 
starts, e.g., when the sender posts the message.
Thus, the ``life cycle'' of $M$ starts at $t$ and ends at $t_v = t+T_v$.
EphPub also involves an encoding function $Encode(\cdot,\cdot)$. Specifically, 
$Encode(M, t_v)$ denotes the encoding of time-bounded message $M$ 
as a function of $M$ itself and the expiration time $t_v$. This is 
the information that is actually exchanged between $S$ and $R$.
The function $Decode(\cdot)$ denotes the inverse operation, i.e., $Decode(Encode(M, t_v))=M$.

\descr{System Requirements.}
We require that a practical and robust Ephemeral Publishing (EphPub) protocol meets the
following minimum requirements. Specifically, it should:

\begin{enumerate}
\item Guarantee {\em retroactive privacy},  
  which we define under two
  different adversarial models (below).
\item Work for synchronous and asynchronous communications. The sender
  and receiver could have intermittent connectivity and do
  not need to be connected at the same time. In particular, the sender
  and receiver could be turned off at anytime in $[t;t_v]$.
\item Rely only on existing primary services, and not on
  yet-to-be deployed services or infrastructure. In addition, to
  avoid single points of failure or trust, it should not
  depend on centralized services or require the existence of specific
  hardware or devices, such as a TPM (Trusted Platform Module). 
%
\end{enumerate}

\descr{Assumptions.} We rely on the following assumptions.  
(i) $S$ and $R$ securely erase the plaintext message $M$ or key
material from their local storage.
(ii) Messages are only stored or transmitted encrypted.
(iii) $S$ and $R$ know the expiration time $t_v$.

\descr{Retroactive Privacy against a Weak Adversary.}
A weak adversary (W-ADV) has access to the same primitives and services as any user $S$ and $R$.
She can inject, alter, and replay any message between $S$ and $R$.  Besides, after message
expiration time, W-ADV may have full access to $S$ and $R$'s internal memory. Further,
{\em we assume that W-ADV does not have access to $Encode(M,t_v)$ before expiration time $t_v$},
i.e., she does not obtain information exchanged between $S$ and $R$.\\
\indent Formally, we say that EphPub is {\em retroactive-private against a weak adversary} if any efficient weak adversary W-ADV can win
the following game with probability non-negligibly over 1/2. The game is between W-ADV and a two-sided challenger $Ch = (Ch_S, Ch_R)$:
\begin{enumerate}
\item[1.] W-ADV announces two equal-length messages $M_0, M_1$, validity starting time $t$ and expiration time $t_v=t+T_v$.
\item[2.] At time $t$, $Ch$ randomly selects a bit $b \in \{0,1\}$, computes $Encode(M_b, t_v)$, and transfers it from $Ch_S$ to $Ch_R$ according to EphPub.
\item[3.] W-ADV may inject, replay, and modify messages in the communication between $Ch_S, Ch_R$.
\item[4.] {\bf After time $t_v$}, W-ADV accesses $Encode(M_b, t_v)$.
\item[5.] W-ADV outputs $b'$ (and wins if $b'=b$). 
\end{enumerate}
An example of such adversarial setting is an investigation authority
wishing to obtain emails previously sent or received by a user,
e.g., Alice.  The authority may obtain a court order to seize Alice's PC, as
well as to subpoena emails stored by Alice's email provider. That is, even
if email messages were encrypted, the authority may obtain the related keys. 
However, the authority is assumed not to (constantly) monitor the communication channel used to exchange
emails.

\descr{Retroactive Privacy against a Strong Adversary.} 
A strong adversary (S-ADV) has
the same capabilities as a weak adversary.  Additionally, S-ADV may
have access to $Encode(M, t_v)$ at any time.  In other words, she may
eavesdrop all information exchanged between $S$ and $R$.\\
\indent Formally, we say that EphPub is {\em retroactive-private against a strong adversary} if any efficient strong adversary S-ADV can win
the following game with probability non-negligibly over 1/2. The game is between S-ADV and a two-sided challenger $Ch = (Ch_S, Ch_R)$:
\begin{enumerate}
\item[1,2.] Same as above.
\item[3.] S-ADV may eavesdrop, inject, replay, and modify messages in
  the communication between $Ch_S, Ch_R$.
\item[4.] {\bf At  any time}, S-ADV accesses $Encode(M_b, t_v)$.
\item[5.] Same as above.
\end{enumerate}
An example of this stronger adversarial setting is a company
concerned with behavior of one of its employees, e.g., Bob. The company
may obtain his emails from the internal mail server and seize Bob's
PC, but it may also log all traffic in corporate network.

\section{The EphPub Protocol}\label{EphPub_Protocol}
In this section, we describe our EphPub construction, which relies on DNS caching.
Our intuition is as follows. A user $S$ wants to send a message
to user $R$ with validity period $T_v$: $S$ encrypts the message using
a key $k$, so that the key is accessible {\em only} within the
validity period. Each bit of the key is \emph{distributed} on
separate DNS cache resolvers. Entries in the DNS cache expire
according to their Time-To-Live (TTL), thus, the encryption key
can no longer be recovered.

\subsection{Building Blocks}
\label{background}
%
%
%
%
%


\descr{DNS Caching.} DNS Caching allows to reduce the load on
individual DNS resolvers~\cite{RFC1035}. After a successful name
resolution (following a DNS query), the DNS resolver keeps the record
in cache for the time specified in the record's {\em Time-To-Live}
(TTL) value (in seconds).  This speeds-up responses to subsequent
queries, since these will be answered directly from the cache, without
any other query.
The TTL is defined by the domain administrator for each authoritative
DNS record.  Typical values of the TTL are from 1 to 5 days, but this
period may vary from seconds to weeks~\cite{RCF1912}. We will present
the results of our own measurements in Section~\ref{experiments}.

\descr{Open DNS Resolvers.} The Internet features a large number of
open DNS resolvers\footnote{A DNS resolver is open if it provides
  resolution for clients outside of its administrative domain. This is
  not to be confused with the OpenDNS company.} -- devices that
respond to DNS queries on port 53. EphPub relies on open DNS resolvers
allowing recursive queries and performing caching. In the rest of the
paper, we denote them as \emph{DNS cache resolvers}.
During our experiments, we collected 900,000 open DNS resolvers' IP
addresses scanning arbitrary address ranges. Our
study 
reveals that more than $10\%$ of the identified open DNS resolvers
perform caching properly. As Dagon et al.~\cite{Dagon08} estimated the
number of open recursive DNS resolvers to $17$ million, we could
estimate the number of open DNS cache resolvers to 1.7 million. For
additional measurement studies on the DNS, we refer
to~\cite{Kristoff,dnssurveys}.

\subsection{Ephemeral Bits}\label{ephemeral_bits}
We now describe how EphPub encodes ephemeral bits: the existence (resp., the absence) 
of a record in a particular resolver's cache is associated to the bit $1$ (resp., $0$). To store an
ephemeral bit $1$, $S$ performs a recursive DNS request of an
existing domain name, $dname$, to a DNS cache resolver, $res$.
The resolver replies with the domain's record and caches the response
for a period of time corresponding to domain's TTL\footnote{To ease exposition, we assume for now that TTL is equal to 
the desired validity period, $T_v$.}. Hence, the existence of an entry in the cache for a given domain name
is interpreted as a bit $1$. Whereas, a non-existing entry is
considered as a bit $0$.
At time $t$, $S$ can transfer the bit $1$ to $R$ by sending the
triplet $\{dname,res,t_v\}$ (being $t_v=t+T_v$).  To store an ephemeral bit $0$, $S$
sends the same triplet to $R$ {\em without} performing any recursive
DNS request.  To read the bit $R$ performs a non-recursive DNS request of the domain name $dname$ to $res$.
If $dname$ is in the cache, $res$ replies with the corresponding
entry and the bit is read as $1$. If the entry is not in the cache the bit
reads as $0$.

If a DNS request is performed once $t_v$ has passed, $res$ will reply
with an empty record, i.e., the bit will read as $0$ independently of
its original value.

\subsection{Protocol Description}\label{protocol_description}
We now extend our approach to implement ephemeral bits to distribute
(resp., retrieve) an entire encryption (resp., decryption) key.

\descr{Message Encryption.} First, the sender selects a random $n$-bit key $k$,
for a semantically secure cipher, such as AES. $k_i$ denotes the $i$-th bit of the key.
Then, the sender encrypts a message $M$, under key $k$, and produces the corresponding ciphertext, $CT$. 

\begin{figure}[t!]
\centering
    \includegraphics[width=0.56\textwidth]{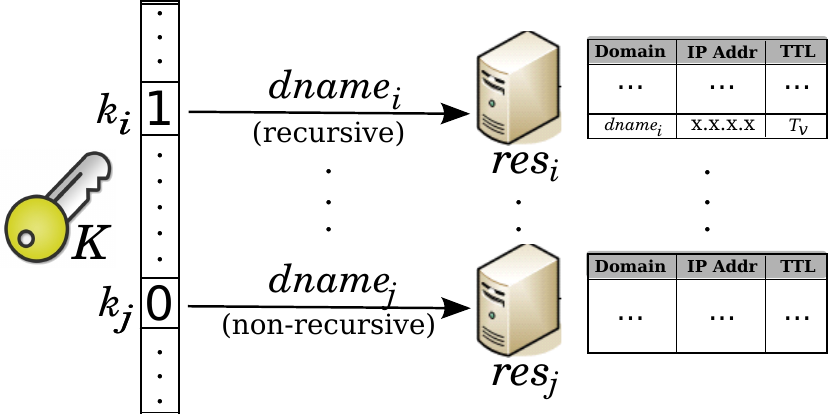}
  \caption{EphPub Key Distribution. \label{fig:write} 
  }
\end{figure}

\descr{Key Distribution.} The distribution of ephemeral key bits over
DNS cache resolvers is illustrated in Figure~\ref{fig:write}.
Note that this instantiates the $Encode(\cdot,\cdot)$ function introduced in Section~\ref{sec:req}.
The sender $S$ generates $Rs=\{res_1,\ldots,res_n\}$, a list of $n$ random (different) DNS cache resolvers, and $Dn=\{dname_1, \ldots, dname_n\}$,
a list of $n$ random valid domain names that have a TTL significantly
close to $T_v$. Each domain $dname_i$ is selected as follows: $S$ first picks a candidate $dname_i$
at random from a (precomputed) list of domains with the related TTL; then, 
it verifies if $dname_i$ is already in the cache of $res_i$, through a non-recursive DNS request of $dname_i$. If $res_i$ replies with a
valid record, $dname_i$ is discarded, otherwise it is added to $Dn$. 

Note that the list of random domains is precomputed and distributed to users. Specifically,  
we generate a random IP address, execute its reverse lookup, and check whether the
corresponding domain name had a satisfactory TTL\footnote{An example of domain names is: softbank126060235192.bbtec.net.
Also, one could use existing  databases of host names, e.g., robtex.com.}. 
Next, for each bit $k_i$ of the key such that $k_i=1$, $S$ performs a
recursive DNS request of the domain name $dname_i$ to the DNS cache resolver
$res_i$. Consequently, the resolver resolves the domain name,
replies to the sender, and populates its cache with the corresponding record. 


\descr{EphPub Object.} The sender $S$ builds the
{\em EphPub Object} (EPO), defined as the output of the $Encode(\cdot,\cdot)$ function, introduced in Section~\ref{sec:req} and instantiated above.
Thus, we define $EPO(M, t_v)=\{CT,Rs,Dn,t_v\}$, i.e., the EPO is composed of the
ciphertext ($CT$), the list of DNS cache resolvers ($Rs$), selected domain names ($Dn$), 
and content expiration time, defined as $t_v=t+T_v$. 
The ciphertext is the result of the encryption of message $M$ using an ephemeral key
whose bits are distributed over DNS cache resolvers $(res_1, \cdots, res_n)$ 
by querying domains $(dname_1, \cdots, dname_n)$, respectively.

\begin{figure}[t!]
\centering
 \includegraphics[width=0.56\textwidth]{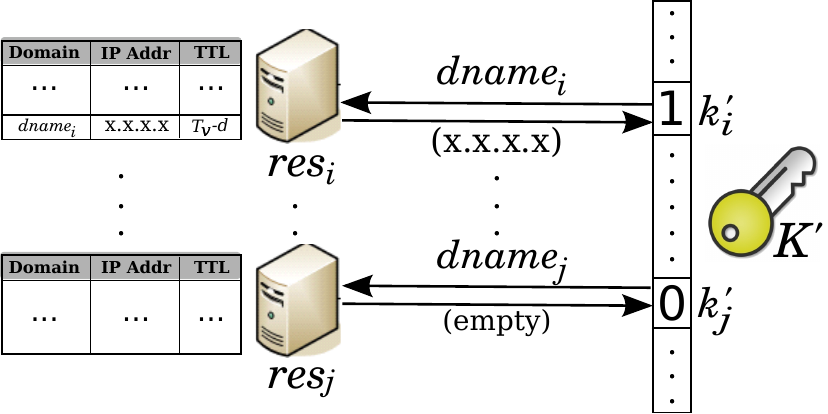}
\caption{EphPub Key Retrieval.\label{fig:read}} 
\end{figure}

\descr{Key Retrieval.} Upon reception of the EPO, i.e., $EPO(M,t_v)=\{CT,Rs,Dn,t_v\}$, the receiver checks whether the current time is smaller than $t_v$. 
If this is the case, she reconstructs the key $k'$ as follows.
The retrieval of key bits from DNS cache resolver (at a time $t+d$, s.t. $t+d<t_v$) is also illustrated in Figure~\ref{fig:read}.
For each domain name $dname_i$, ($i\in{1,n}$), the receiver performs a
non-recursive request to the DNS cache resolver $res_i$. If $res_i$
replies with a valid record, then $k'_i$ is set to 1, otherwise, $k'_i$ is set to 0.
Once the key is retrieved, she decrypts 
the ciphertext $CT$ to retrieve the message $M$. These operations instantiate the $Decode(\cdot)$ function introduced in Section~\ref{sec:req}.

\smallskip\noindent Note that the message
is \textbf{\em never} transmitted or stored in cleartext, but only in its
encrypted form, specifically, in its ``EPO'' form.  This guarantees
that the message $M$ cannot even be retrieved with a forensics analysis on
the hosts after $t_v$. 
Remark that, after the message validity time $t_v$, the DNS records will be removed from
the cache of the DNS resolvers listed in $Rs$. Thus, the key will ``disappear'' and the ciphertext can no longer be decrypted.

\subsection{EphPub+: Extending EphPub Against a Strong Adversary}
\label{sec:ext}
In case the EPO is transmitted over a communication channel, 
the confidentiality of the channel becomes relevant.
Consider the case of the strong adversary presented in
Section~\ref{sec:req}: such an adversary may eavesdrop on the
communication channel, get hold of the EPO before expiration time,
and hence retrieve the ephemeral key and the message $M$.
We extend the EphPub protocol above to be robust even in presence of a strong adversary.
Besides executing the EphPub protocol, the sender super-encrypts the EPO, using, for instance,
an asymmetric encryption scheme, such as PGP, and the receiver's long term public key. Note that, if there
are several receivers, broadcast encryption techniques~\cite{broadcast} can be used to minimize
related overhead.
In the rest of the paper, we will refer to EphPub+ as the protocol enforcing super-encryption of the EPO prior to
transmission, and we refer to {\bf $\mathcal{E}(EPO)$} as the encryption of the EPO.

\descr{Remark.}
 One additional potential concern with
  respect to a strong adversary is related to monitoring (cleartext)
  DNS messages between the sender and the cache resolvers.  Because 
  of this, queries to the DNS resolver have to be performed in
  random order. Thus, assuming $128$-bit keys, a strong adversary
  monitoring a server's DNS traffic learns the Hamming weight of the key
  (i.e. the number of ones), which corresponds to reducing the
  entropy of the key by roughly $5$ bits\footnote{In average~\cite{Messerges:2002:ESS:570513.570522}, 
    the key search space for a key of size m with known Hamming weight
    is of $\sum_{m=0}^n \frac{ \binom{n}{m}^2}{2^n}$.}.




\modified{If this marginal reduction of the key strength is 
problematic, slightly increasing the key allows to compensate for it.
For example, using a key of $134$-bits provides
slightly more than $128$-bits of security when the Hamming weight of the
key is known to the adversary}.  Alternatively, the sender can tunnel
DNS requests, e.g., using Tor~\cite{TOR}. Note that, as a positive
side effect, the use of Tor provides higher anonymity, as DNS requests
cannot be correlated to senders.

\subsection{Handling Errors and Erasures}\label{handling_errors}
Bit {\em errors} may occur in the ephemeral key: for instance,
if a resolver is re-started while its cache is populated with one record used 
by EphPub, the corresponding bit will be decoded as a $0$, instead of a $1$. 
A bit can also flip from $0$ to $1$ if the corresponding record
is requested by another user, after the sender performed key distribution.
Note that the probability of this event is negligible 
if the domain name is selected at random, among a virtually unlimited number 
of host names, e.g., if wild-card domains are used.
Finally, \emph{erasures} occurs when the corresponding bit cannot be
set, for example when a resolver does not reply during key retrieval.

To address potential errors and erasures, we use error correction
codes~\cite{mackay2003information}, which make data resilient
to a defined number of errors.
%
Specifically, we use Reed-Solomon (RS) codes~\cite{reed-solomon}. We select a $(63,55)$ RS code allowing to
correct between 8 and 48 erasures or between 4 and 24 errors (depending on
where the errors occur, since symbols are 6 bits long) or a combination of those, within the capacity of the error correction code. 
The use of this code in EphPub increases the number of DNS requests only by $20\%$, e.g., for a $128$-bit key, 
the increase is from 128 to 176. 
Note that the receiver needs to fetch the \emph{correction bits} only when errors or
erasures are present.
Our experiments show that the Reed-Solomon code guarantees accurate key decoding
with very limited overhead.  If needed, error correction capabilities can be tuned 
by changing the parameters of the code, allowing higher reliability or more efficient network usage.

\section{Security Analysis}\label{security_analysis}
In this section, we analyze the security and privacy provided by
EphPub. 
We also discuss several attacks that may target the EphPub protocol as well as potential issues that might be used to compromise system's
security.

\subsection{Retroactive Privacy in EphPub}
\noindent{\bf Theorem 1.} {\em EphPub is a retroactive-private in presence of
a weak adversary}.

\vspace{0.2cm}
\noindent{\bf Proof }({\em Intuition}).
We observe that no efficient weak adversary, W-ADV, targeting an EphPub execution, has a non-negligible advantage over 1/2 
in the game introduced in Section~\ref{sec:req}.
%
First, remark that, as the weak adversary does not get any additional information during $[t,t_v]$, the only advantage for her may derive from accessing 
$EPO(M_b, t_v)$, at time $\tau > t_v$, i.e., after the EPO has expired. 
Intuitively, an adversary accessing an expired EPO retrieves
the ciphertext of the target message, the list of DNS cache resolvers
with the related queried domain names. Assuming the underlying encryption
scheme is semantically secure, the ciphertext does not reveal any additional information, unless
the corresponding key is retrieved. 
Recall that EphPub uses the expiration of DNS cache records to encode the key:
this inherent property of DNS resolvers causes significant challenges to
attackers that do no target their attacks prior to data expiration.
Indeed, although the EPO still contains the list of DNS cache resolvers that 
were used for key storage, these resolvers no longer contain key
information. In fact, even if resolvers are subpoenaed or subject to
forensics, related key bits would not be present in the cache. 
Thus, after the TTL has passed, attackers cannot learn whether the domain
name has ever been in the cache.
Also, observe that, since the receiver erases the encryption key and
the message $M$ before its expiration time, potential compromise  of the receiver does not
help the attacker. 

As a result, the adversary cannot use information obtained from an
expired EPO to recover the original message, 
thus, EphPub is {\em retroactive-private}.

\subsection{Retroactive Privacy in EphPub+}
\noindent{\bf Theorem 2.} {\em EphPub+ is a retroactive-private in presence of
a strong adversary.}

\vspace{0.2cm}
\noindent{\bf Proof }({\em Intuition}).
We remark that no efficient strong adversary, S-ADV, targeting an EphPub+ execution, has a non-negligible advantage over 1/2 in the game introduced in Section~\ref{sec:req}.
%
%
Recall that S-ADV has access to the communication between $Ch_S$ and $Ch_R$, as opposed to W-ADV. Thus,
the advantage for S-ADV may derive both from accessing $EPO(M_b, T_v)$ at time $\tau > t+T_v$ and 
from accessing $\mathcal{E}(EPO(M_b, T_v))$ at any time before $t+T_v$.
However, one can reduce the retroactive privacy in presence of S-ADV to the security of the encryption scheme $\mathcal{E}$ {\em and} the 
retroactive privacy in presence of W-ADV. Indeed, $\mathcal{E}(EPO(M_b, T_v))$ is a ciphertext produced using a secure encryption scheme 
(as discussed in Section~\ref{sec:ext}), hence, it is straightforward to show that, if S-ADV has a non-negligible advantage in distinguishing $M_b$ given the knowledge of $\mathcal{E}(EPO(M_b, T_v))$, then S-ADV can be used to break the security of the underlying encryption scheme. 
In fact, since the adversary's view is again restricted to $EPO(M_b, T_v)$ \emph{after} expiration time $t$+$T_v$, we can use the same arguments
of Theorem 1. 

\subsection{Infeasibility of Key Recovery Attacks}

\descr{Brute Force Attack.} In order to reconstruct an EphPub key, an
attacker may attempt to crawl all existing DNS cache resolvers with 
DNS requests, and try to identify some bits of the
key. However, since the number of DNS cache resolvers and, especially,
that of all possible domain names is extremely large, this approach is
not feasible. Note that, in early 2011,
  \url{http://www.whois.sc/internet-statistics} reported $~126$
  million active registered domain names. Moreover, relying on a
  sufficiently large set of wild-card domains would virtually remove
  any limit.

\descr{DNS Infrastructure Infiltration and Sybil Attacks.}  An additional
adversarial strategy could involve infiltration in the DNS infrastructure, 
somehow similar to the Sybil attack, performed against the DHT used by
Vanish~\cite{Defeat-Vanish}. Recall that, in a Sybil attack, an attacker controls
several hosts that generate many virtual identities, hence, she receives a
very large portion of the traffic.  However, DNS resolvers are
uniquely identified by their IP addresses and do not have virtual
identities.  A Sybil attack or an infiltration attack on the DNS
infrastructure would require either (i) a very large number of public
IP addresses pointing to a few hosts, (ii) a very large number of
hosts acting as DNS resolvers or (iii) compromising or obtaining
traffic logs of a large amount of the DNS resolvers.

We argue that those options are far from being viable.  
In fact, unless the attacker controls an extremely significant portion of 
DNS resolvers used to store the key of the target EPO, the amount of information
that the attacker could learn is very limited. Since
resolvers are chosen at random, e.g., among a million servers,
an attacker that compromise a fraction of those would obtain, on average,
the same fraction of key bits. That is, controlling
$100K$ resolvers would lead to recovering only $10\%$ of key bits.

Also note that an attacker cannot use newly
deployed resolvers or compromised bots, which are often unstable and
online for a relatively short period of time. In fact, in the design of 
EphPub, we make sure that resolvers are picked from a list that is
generated and maintained conservatively (see Section~\ref{experiments}).

\subsection{Denial of Service}\label{dos_analysis}
We now consider potential DoS attacks against EphPub. We identify two scenarios. In the first one, the attacker knows
the EPO and her goal is to prevent the receiver from recovering the message.
In the second scenario, the attacker is not focusing on a specific target,
whereas, her goal is to attack the entire EphPub service.
 
\descr{Attacking a known EPO.} An attacker with access to an EPO has
a couple of potential approaches to prevent decryption. First, all key
bits could be flipped to 1 by performing recursive DNS requests of all
the EPO domain names. However, after this attack the records
corresponding to bit $0$ in the initial key will have a larger TTL
than the records corresponding to bit $1$. Therefore, the initial key
could still be recoverable by the receiver. Second, the attacker could
target the caches of DNS resolvers, which have finite memory.
As a result, an attacker could launch a DoS to EphPub by
filling caches with random entries until the entry used to store a bit
is removed from the cache.  However, the default Bind behavior does
not limit the cache size, making this attack almost impossible.  For DNS
cache resolvers with cache size limitations\footnote{Note that BIND9 has
an option to limit cache size \texttt{max-cache-size}, however it
has been reported to be nonfunctional.} the attack would be
feasible at the cost of a huge number of DNS queries.  Finally, the attacker could perform a DoS attack on the DNS
cache resolvers themselves, by flooding them with bogus DNS requests.
Note that as for most of Internet protocols, DoS (or even worse, DDoS) attacks
are very difficult to prevent and are in general unrelated to the
privacy of the system. One natural countermeasure is to always enforce EphPub+, 
i.e., super-encrypting the EPO, so that an attacker would not have access to the DNS information.


\subsection{Traffic to Top Level Domains and Root Servers}
\label{sec:top-level-domains}

When a bit is stored in a DNS cache resolver, the resolver will need to
walk the DNS hierarchy to resolve the hostname's components it does not
currently have in cache. Assuming a completely empty cache, the
following steps will be preformed by the DNS cache resolver to resolve
the hostname ``\emph{sub}.\emph{domain}.\emph{tld}'':

\begin{itemize}
\item A root server $R_i$, among {$R_1, \ldots, R_n$}, is queried for
  \emph{sub}.\emph{domain}.\emph{tld} $R_i$'s will answer the
  addresses of the authoritative servers for \emph{tld} (Top Level
  Domain servers), e.g., {$T_1,\ldots,T_n$}.
\item A TLD server $T_j$ is queried for the record
  \emph{sub}.\emph{domain}.\emph{tld}. It will answer the list of
  authoritative servers for the domain \emph{domain}.\emph{tld},
  {$A_1,\ldots,A_n$}
\item Finally, an $A_k$ will be queried for the domain
  \emph{sub}.\emph{domain}.\emph{tld.}, and will answer with the
  requested record or a negative answer if this record does not exist
  (or a list of DNS resolvers in case there is further sub domain
  delegations).
\end{itemize}

During the above process, the key might be disclosed to $R_i$, $T_j$,
even if they are not malicious themselves, their network traffic might
be logged, e.g., for forensics purposes.  If an attacker gets access
to the logs of all root and/or all TLD servers involved, she can then
retroactively recover the original message from an EPO.

An effective counter-measure is to ``pre-fetch'' TLD and domain
records, independently of the value of the bit to store, by performing a
recursive query for an arbitrary record of the same domain, e.g.,
\emph{www}.\emph{domain}.\emph{tld}. This will force the cache
resolver to obtain the records for \emph{tld} and \emph{domain.tld}.
Therefore, this will only exposes a hostname independent of the
hostname actually used to store EphPub bits. Also, the information
received by root and TLD servers is the same, independently of the
value actually stored in the DNS cache.
Our prototype implements ``{\em pre-fetching}'' by default, this has a
moderate impact on the performance and overall network load, as it
performs requests independently of the value of the bit to store.

%

\subsection{Irrelevance of  Known DNS Security Issues }\label{misc}

\descr{DNS Flushing Periods.} Older versions of the BIND DNS Server do not periodically erase
expired entries from the DNS cache. 
Therefore, a forensic analysis could allow to recover the entries (i.e., key
bits), after their normal expiration date.  This analysis would
 reveal parts of a key that should not otherwise be available. Fortunately, only BIND versions older than 4.9 lack a regular
cache cleanup~\cite{oreilly_bind}. In our experiments, there were only $4$
out of $90K$ resolvers which were fingerprinted as using BIND versions before 4.9, thus,
we excluded them.  
More recent BIND versions define a 1-hour default cleaning interval. Expired
entries encoding the key bits are therefore removed from the cache
with at most one hour delay.

\descr{DNS Poisoning Attack.} The DNS system has been the target of well-known
attacks, such as DNS cache poisoning~\cite{BIND_cache_poison}. In such an attack, an
attacker is able to inject a fake record associating a legitimate host
name to a bogus IP address. This attack is used for redirecting users
to a malicious site. Variants of this attack
rely on injecting fraudulent glue records together with normal
response~\cite{cert_Kaminsky} in order to redirect queries to
malicious resolvers. This attack does not interfere with EphPub as
long as records can be cached by resolvers, even if the stored
values point to malicious IP addresses.

\descr{Domain Name System Security Extensions.} DNSSEC have been
recently designed to provide authentication and data integrity of DNS
records~\cite{DNSSEC}.  The main idea of DNSSEC is to digitally sign
answers to DNS lookups using public-key cryptography.  EphPub is not
affected by the deployment of DNSSEC since these extensions do not
impact the DNS caching mechanisms EphPub is built on.

\descr{DNS Proxying.} A few enterprise
networks and ISPs implement transparent DNS proxying~\cite{RFC5625} to
reduce bandwidth overhead.  While not recommended, DNS transparent
proxying forces a sender to use the DNS resolver chosen by the ISP. The
DNS proxy intercepts any incoming DNS request and redirect them to the
configured resolver.  As a result, a sender cannot select the DNS resolver
to resolve her request and, therefore, EphPub cannot operate.  A
similar problem arises when full packet capture and storage is
performed within a close network distance to the sender or receiver. In
this case, captured packets, together with the EPO, can result into
recovering the message.  However, these problems can be addressed by
tunneling  EphPub's DNS requests, e.g., using Tor~\cite{TOR}. 

\section{The EphPub Prototype}\label{software}
 We have implemented an EphPub prototype with a simple modular architecture made of two components:
 the ``DNS core'' (backend) and the user interface (frontend).
 In this section, we present the details of the components and the interface between them, as well as
 the description of our software implementations -- the source code of our
  implementations is available at \url{http://code.google.com/p/ephpub/}. 

\descr{Backend.} The crucial component in EphPub is the software layer responsible for performing DNS queries to distribute/retrieve key bits.
 In the EphPub prototype, this layer has been implemented in Python, enhanced by
 the open-source PyDNS module (\url{http://pydns.sourceforge.net}) 
 for DNS queries.
 Given the simplicity of DNS queries, EphPub can easily be ported
 to any other system/architecture. 
 To distribute ephemeral keys, EphPub uses randomly picked DNS cache resolvers
 and domain names.
 The implementation of key storage is straightforward, by means of  recursive and non-recursive DNS queries.
 As a result, the footprint of our core implementation is extremely compact.
 Encryption is done with AES, relying on the PyCrypto python module (\url{http://www.dlitz.net/software/pycrypto}) and 128-bit keys.

 \descr{Frontend.} 
 Given the simple structure of our Python-based core, the development
 of the user interface is not limited by particular assumptions that
 may affect portability or usability.
 Besides a simple {\em command-line tool} for ephemeral files, we
 developed a {\em Firefox anf Thunderbird extension} prototype, using the python
 support provided by the {\em pyxpcomext} extension
 (\url{http://pyxpcomext.mozdev.org}). Therefore, all 
 DNS operations are performed from within the extension.  This
 extension allows: (1) the sender to encrypt any web content 
 using EphPub and publish the corresponding EPO, (2) the receiver to
 open an EPO and decrypt the corresponding information (as long as that
 expiration time is not passed).

 \descr{Usability.} 
 The EphPub functionality is completely included in a stand-alone Firefox and Thunderbird extension,  i.e., we do not require
 the user to  install or launch any additional background software, as opposed to Vanish~\cite{Vanish}.
 We believe that this is crucial in order to improve the usability of the solution. Specifically, the Vanish Firefox extension
 requires an underlying DHT backend (i.e., Vuze) that needs to be launched  independently. This also involves
 potentially ``invasive'' operations, such as the installation of the Java Virtual Machine. 
 In contrast, our prototype only requires installing the Firefox or Thunderbird extension,  thus, it addresses a larger variety of potential users and devices.

 \descr{Availability.} Our prototype does not incur a setup delay, as opposed to Vanish,
 where the use of the extension is subject to the Vuze backend to bootstrap. 
 This can take 5-10 minutes according to~\cite{Vanish}, and it may increase, as we have
 experienced, depending on local network configuration.
 Furthermore, we stress that P2P-based solutions may be prevented by
 firewalls and network filters, whereas, EphPub uses simple DNS queries, which are unlikely to be
 firewalled. 



\section{Measurements \& Experiments}\label{experiments}
We now overview our measurements, performed over several months, to select DNS cache resolvers,
and present our experimental analysis of EphPub's efficiency and robustness.

\subsection{DNS Cache Resolvers}\label{exp_cache}
EphPub uses a set of randomly chosen DNS cache resolvers to distribute encryption keys.
Thus, the related software layer must be equipped with an appropriate list.
Although EphPub addresses potential errors and failures using error correcting codes, 
we need to assess DNS resolvers' reliability to minimize error rate.
To this end, we build a list containing the highest possible ratio of resolvers that:
(i) successfully respond to both recursive and non-recursive queries, (ii) have a small rate of connection errors,
(iii) maintain entries in the cache for the entire TTL time and not longer.
Since an accurate list of such resolvers would evolve over time, we design EphPub to mirror this evolution
by providing users with automatic periodical updates. 

We now overview our methodology. We remark that it might be worth investigating how to maintain an optimal dataset, e.g.,
drawing from existing studies, such as~\cite{Dagon08,Kristoff,dnssurveys}.

\descr{Initial DNS Resolvers Dataset.}  We started from a list of 900$K$ IP addresses responding to DNS queries,
from a previous probing performed several months earlier.
We re-probed these addresses and obtained a list of $\sim$225$K$ active IP addresses.
We validated this list by verifying whether the addresses answered to DNS queries throughout a period of 2 months and
filtered those addresses that:
(i) did not answer to recursive DNS queries,
(ii) did not cache answers to recursive queries,
(iii) did not answer to non-recursive DNS queries,
(iv) performed caching or recursive resolution when receiving non-recursive queries.
This way, we built a dataset of $\sim$130$K$ DNS cache resolvers.

\descr{TTL Compliance.} Next, we focused on selecting cache resolvers that register the correct TTL. Indeed, looking at the TTL of the resolution answer returned after a recursive query,  we noticed that some resolvers ignore the domains' TTL, as already pointed out by~\cite{Dagon08}.
After a 4-week observation, we reduced our dataset to $\sim$80$K$ resolvers to exclude misbehaving resolvers.
Finally, we made sure that DNS cache resolvers used by EphPub would  successfully {\em maintain} cache entries for the intended time, i.e., domain's TTL.
Hence, we periodically performed recursive DNS queries and subsequently test the presence of the corresponding cache entry
with non-recursive queries. 
Following a conservative strategy, we discarded resolvers that generated failures or premature cache entry removals after an observation time of 4 weeks. 
As a result, our final dataset includes $\sim$25$K$ {\em reliable} resolvers.
We summarize our measurements in Table \ref{table:dnscacheresolvers}. 

\begin{table}[h!]
\begin{center}
{\footnotesize
\begin{tabular}{|c|c|c|c|}
\hline
{\bf Experiment} & {\bf Dataset} & {\bf Fail} & {\bf Pass} \\
\hline
Initial dataset & {900$K$} & {675$K$}& {225$K$} \\
\hline
Correctly perform cache & {225$K$} & {95$K$}& {130$K$} \\
\hline
Cache correct TTL &  {130$K$} &  {50$K$} &  {80$K$} \\
\hline
Cache persistence & {90$K$} & {65$K$} & {25$K$} \\
\hline
\end{tabular}
}
\caption{Building a dataset of reliable DNS cache resolvers.\label{table:dnscacheresolvers}}
\end{center}
\vspace{-0.4cm}
\end{table}

\subsection{Domain Names}\label{exp_domains}
One of our requirements is to let users actively and accurately control message expiration time.
In EphPub, this time is related to key expiration time, thus, to the cache entries' TTL.
To this end, EphPub performs recursive queries only on domains that have the desired TTL.
Recall that the set of domains is generated at random, as described in Section~\ref{protocol_description}. Although the RFC1912~\cite{RCF1912} defines the TTL to potentially last up to several weeks, in practice, we find that only a negligible fraction of domains have TTL higher than $604,800$ seconds (1 week).
We highlight that the range of possible TTLs a user can select is
reasonably wide. To confirm this, we generated 2 million random
domains that could be successfully resolved (i.e., we generated a
random IP address and performed a reverse lookup to obtain the
corresponding domain name). Then, we harvested their TTL provided by
the authoritative DNS server. We notice that TTLs range from 0 seconds
to 7 days.  The distribution of most frequent TTLs is shown in Table
\ref{tab:TTLdistr}.  As a result, our current techniques limit
expiration time to 1 week.  In many cases such a lifetime
is enough for the application we consider (e.g., the {\em Online
 Social Networks} scenario in Section~\ref{Introduction}).  Also, note that this is a significantly larger
interval than prior work: in Vanish~\cite{Vanish}, key shares are
deleted after (typically) 8 hours, due to DHT churn.  However, we
acknowledge that it is an interesting open challenge how to support
even longer expiration times, e.g., by periodically (and
automatically) re-initializing ephemeral keys.\footnote{In cases where longer lifetimes are needed, it is possible
    (although not very efficient) for the source of the content to extend
    the expiration time of the DNS entries, by refreshing the cache
    after expiration, or to create a new EPO and distribute the new
    EPO to users.}

\begin{table}[h]
\footnotesize
\begin{center}
\begin{tabular}{|l|r|l|l|r|}
\cline{1-1}\cline{2-2}\cline{4-4}\cline{5-5}
\multicolumn{1}{|c|}{\textbf{TTL}} & \multicolumn{1}{c|}{\textbf{domains}} &  & \multicolumn{1}{c|}{\textbf{TTL }} & \multicolumn{1}{c|}{\textbf{domains}} \\\cline{1-1}\cline{2-2}\cline{4-4}\cline{5-5}\cline{1-1}\cline{2-2}\cline{4-4}\cline{5-5}
1200 [20m]     & 13,595    &  &86400 [24h]    & 998,450   \\ \cline{1-1}\cline{2-2}\cline{4-4}\cline{5-5}
1800 [30m]     & 7,269    &  & 172800 [2days] & 77,326   \\ \cline{1-1}\cline{2-2}\cline{4-4}\cline{5-5}
3600 [1h]      & 201,789   &  &259200 [3days] & 12,317    \\ \cline{1-1}\cline{2-2}\cline{4-4}\cline{5-5}
7200  [2h]     &171,685    &  &432000 [5days] & 13,450    \\ \cline{1-1}\cline{2-2}\cline{4-4}\cline{5-5}
43200 [12h]    & 180,144   &  &604800 [7days] & 42,142    \\ \cline{1-1}\cline{2-2}\cline{4-4}\cline{5-5}
\end{tabular}
\end{center}
\vspace{-0.3cm}
\caption{Number of most frequent occurrences of TTLs over 2 million random domains.}
\label{tab:TTLdistr}
\vspace*{-0.4cm}
\end{table}

\subsection{EphPub Robustness}\label{robustness}
We now evaluate the robustness of EphPub system: we test
distribution and retrieval of EphPub keys.  Our goal is to examine
whether the keys can be successfully retrieved before the expiration
time while disappearing afterwards.  

To this end, we executed the EphPub protocol introduced in
Section~\ref{protocol_description}. We fixed the encryption key size to
$128$ bits and chose a desired expiration time, i.e., 24 hours and 7
days.  We stored the key as follows. First, we encode the key using
the Reed-Solomon code discussed in Section~\ref{handling_errors},
obtaining a $176$-bit encoding, denoted with $e_1, \cdots, e_{176}$.
Next, for each $e_i$, we wrote $e_i$ on the randomly picked cache
resolver $res_i$ using the domain name $d_i$ chosen at random and having a
TTL equal to the desired expiration time.
Then, at periodic intervals, we retrieved the keys from the cache
resolvers.  For each $(res_{i}, d_{i})$, we performed a non-recursive DNS
query to cache resolver $res_{i}$ for the domain $d_{i}$: if an entry in
the cache existed, we read $e'_{i}=1$, otherwise $e'_{i}=0$.  Finally,
we decoded the key using the Reed-Solomon decoding.

We now analyze the percentage of key bits successfully retrieved
during our experiments.  We measured the correctness of both:
\begin{enumerate}
\item The bits that were read/written on cache resolvers, i.e., whether
  $e'_{i}=e_{i}$.
\item The key bits effectively used for encryption/decryption, 
  after applying error correction code. 
\end{enumerate}

\begin{figure}[t]
\begin{center}
  \subfigure[]{\label{fig:24h}\includegraphics[width=0.4\textwidth]{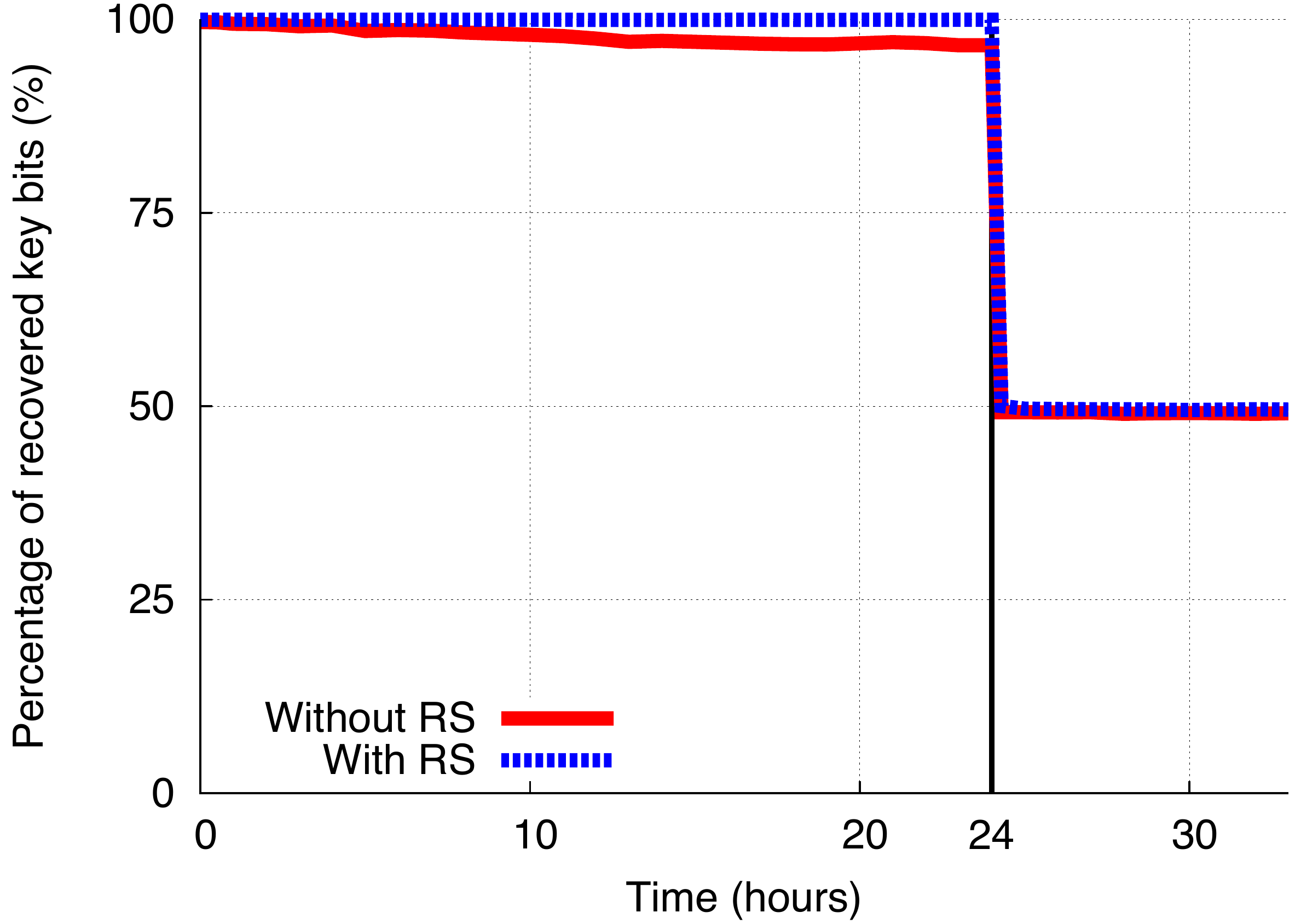}}
  \subfigure[]{\label{fig:1w}\includegraphics[width=0.4\textwidth]{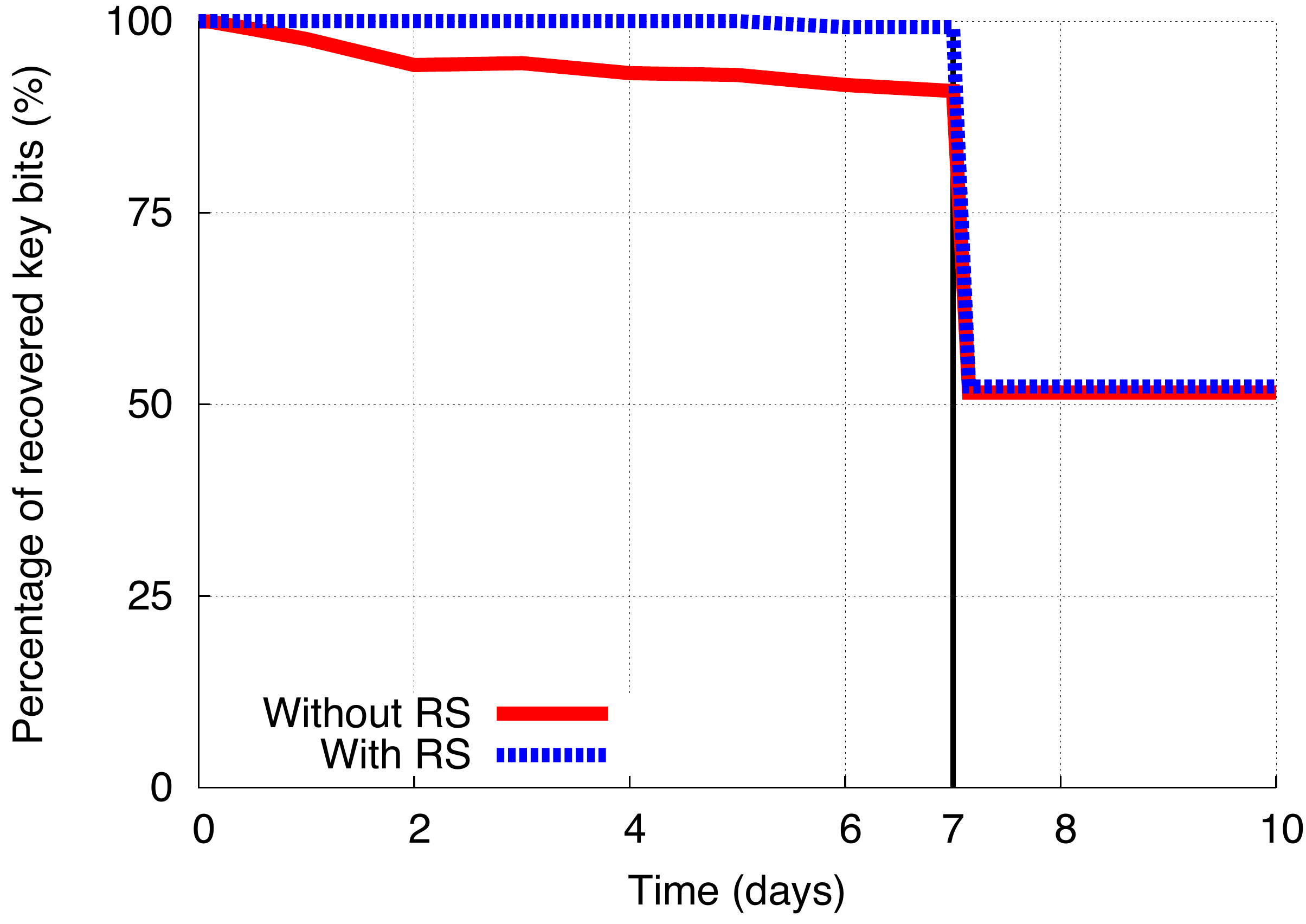}}
  \end{center}
  \vspace{-0.35cm}
  \caption{{Percentage of EphPub key bits correctly recovered with
    expiration times, respectively, 
24 hours~\subref{fig:24h} and 7
    days~\subref{fig:1w}. {\em Before expiration time}, key bits are
    correctly recovered. {\em After expiration time}, key bits
    $1$ (on average 50\% of key bits) flip to $0$ and are no longer
    correctly recovered. Whereas, key bits $0$ (the other 50\% of key
    bits) conserve their value. The resulting key is a dummy sequence
    of 0's and can no longer be used for valid decryption. Note that
    lines are plotted as the interpolation over periodical key
    retrievals.}}
    \label{fig:robustness}
\end{figure}

Figure~\ref{fig:robustness} presents the results of our
experiments (averaged over 100 trials). 
Keys could be correctly retrieved up to the expiration time, whereas
right after TTL timeout all bit 1's flip to 0 in the corresponding
key.  At this point, the recovered key is all 0's (it reads on
Figure~\ref{fig:robustness} as around 50\% of the initial random key
bits, which were initially 0's, are correctly recovered). Clearly,
this provides no information on the original key.

One potential concern regarding EphPub's robustness may be related to
its resilience to errors due to collisions in writing ephemeral
bits. However, one can make collision nearly impossible using
wild-cards. Nonetheless, one can estimate the probability of a
collision (without using wild-cards), using the birthday
problem. Specifically, we model the probability of having at least one
collision as $p(n;d)=1-e^{-n(n-1)/2d}$, where $n$ is the number of
EphPub documents with same expiration time and generated within the
same time frame, and $d$ is the product between the size of cache
resolvers' dataset (i.e., 25$K$ in our experiments) and the number of
domain names with the specific TTL.  For instance, considering
documents with 24-hour expiration time, using our dataset of 25$K$
cache resolvers and 1$M$ random domains with 24-hour TTL, and assuming
$n$=10$K$, we obtain $p(n;d)\approx 10^{-3}$.  Also, recall that our
error correction code can correct between 4 and 24 errors (see
Section~\ref{handling_errors}). These numbers correspond to
  a small to medium deployment. Scaling to a larger number of users
  would require more DNS servers\footnote{\modified{Based on work
      from Dagon et. al.~\cite{Dagon08} we estimate that 1,7 million
      DNS servers on the Internet would be suitable}} and the use of 
wildcard DNS domains to completely avoid the birthday paradox problem.

\renewcommand{\baselinestretch}{0.95}
\renewcommand{\baselinestretch}{1.0}

We conclude that the EphPub prototype is robust and accurate enough for real-world deployment.
Nonetheless, a more accurate and extensive profiling of the DNS cache resolvers may further reduce the
percentage of bit flipping. 

\subsection{Performance Evaluation}\label{performance}
We now analyze the performance of EphPub to confirm that our approach is
fast enough to be used in practice. In addition, we discuss several ideas
to optimize our implementation. Measurements are done on a Dell PC with two quad-core CPUs Intel Xeon at 1.60GHz with 4GB RAM,
Python 2.6.2 (with pydns-2.3.3 and pycrypto-2.1.0b1) and a high-speed Internet connection. 

\descr{Key Distribution.} The main operations involved in distributing EphPub keys are the 
generation of valid random domains with intended TTL and DNS (non) recursive queries.
(We do not consider cryptographic operations, since symmetric key encryption is relatively fast compared to the former tasks).
Random domain generation is performed as follows: we generate random IP address and perform
 a reverse lookup on the local DNS resolver to check whether the corresponding domain name
resolves and has the desired TTL. Hence, this operation involves a number of trials that is directly related to the \emph{popularity} of the desired expiration time
as domain TTL. Due to its limited overhead,
it is possible to execute multiple instances of this operation in parallel.
However, as described in Section~\ref{protocol_description}, the random domain generation is performed ahead of time, and not online.
Nonetheless, we measure the overhead that users would experience if they want to independently generate the domains at run-time.
Due to potential errors or failures, it is advisable to pre-generate a number of random domains strictly larger---e.g., twice as many---
than the bits to store.
In order to generate $320$ random domains (that is almost twice the number of bits to store, when using 128-bit encryption keys
and Reed-Solomon codes), our prototype requires around $5s$ 
for TTL equal to 86,400 seconds (1 day).
We also measure times for less popular TTLs. To generate the same number of random domains but with TTL equal to 14,400 seconds (4 hours), the prototype
requires around 1 minute.\\ 
\indent On the other hand, the time overhead needed to store key bits in DNS caches mostly depends on the delay in executing recursive queries.
Our experiments show that, executing queries in parallel
(instantiating up to 64 threads), this operation takes about 6
seconds.  Although, DNS queries are not computationally expensive (in our
experiments, the CPU usage never went over 2\% throughout the whole
execution) we let the number of parallel threads be a
parameter that can be set by the user according to its needs.

\descr{Key Retrieval.} Time to retrieve the key bits from DNS caches depends on the
the delay of executing non-recursive queries, which is much smaller compared to recursive queries.
In our experiments, it takes only about $1s$ 
 to retrieve 176 key bits. 

\descr{Additional DNS traffic.} One possible objection to the EphPub
system could be that it generates additional traffic towards DNS
resolvers. 
However, the generated traffic is extremely limited in
comparison to the number of DNS queries generated through typical web
navigation.  
To illustrate this, we note as examples that opening the
Firefox default homepage on Ubuntu leads to more than 50 DNS
queries. Furthermore, opening the \url{www.cnn.com} web page generates
about 140 DNS queries.  The number of queries is related to several
factors, such as pre-fetching performed by Firefox, the number of advertisement links, 
and multiple lookups for
each domain (one for each domain in the search list~\cite{man_resolv},
IPv6 and IPv4 requests, etc.).  When compared to these very common
actions, the network and DNS usage of EphPub can be considered
negligible. 
Moreover, performing those DNS requests in parallel (up to 64 parallel
threads in our experiments) might appear as an \emph{aggressive}
behavior. However, each DNS cache resolver used will receive at most two DNS requests, 
which is a negligible workload.  Also, the
generated network traffic is quite limited: we measured that storing/retrieving a key
on the resolvers generates at most 32KB of DNS traffic, corresponding to
the worst case of having all bits set to 1 (given that an average request/response amounts to 180 bytes). 

\descr{Message Overhead.} The information contained in EPOs naturally generates a message overhead. Indeed, beyond the ciphertext, assuming 
the use of a 128-bit key and the Reed-Solomon error correction technique (see Section~\ref{handling_errors}), 
the EPO contains 176 IP addresses for the DNS cache resolvers (each requiring 4 bytes) and 176 domain names (each requiring approximately 20 bytes).
Thus, we estimate the average size of the additional information in the EPO at 4KB. 
According to the measurement work presented in~\cite{IMC04}, this overhead would be equal to less than one tenth of the average email size --
approximately 60KB.
Nonetheless, as this information is textual, the use of compression algorithms would remarkably reduce related overhead.

\section{Related Work}\label{related}

\descr{Forward Secret Privacy.} {\em Forward Secret Privacy (PFS)}~\cite{DiffieOorWie} is the property ensuring that a session key derived from a pair of long-term public and private keys will not be compromised, if one of the long-term private keys is compromised in the future.
An authenticated Diffie-Hellman (DH) key exchange protocol provides this property since the compromise of the long term key does not provide any information about the DH components that were used. However, a PFS system does not guarantee retroactive privacy. In fact, messages are not time-bounded, hence, the receiver or any party obtaining the key can always decrypt messages without any restriction of time. 

\descr{Ephemerizer.}  The introduction of a trusted third party for
erasing data on behalf of users has been proposed by {\em Ephemerizer}
solutions~\cite{ephemerizer07,PerlmanSISW,Perlman05,FADE}. Users
delegate deletion of protected content to third parties (dedicated
servers) that destroy the data or the encryption keys after a
specified timeout. Again, this violates one of our main goals, i.e.,
to avoid the introduction of one or more trustworthy, always-on, and
centralized services that might be compromised.  For instance, recall
the case of the Hushmail email encryption service, which was secretly
providing the cleartext content of encrypted messages to the
U.S. government~\cite{Hushmail}.


\descr{Vanish.}  Similar to Vanish~\cite{Vanish}, EphPub aims at providing retroactive privacy, even
if data storage is not trusted, compromised or stolen. However, as
discussed in Section~\ref{Introduction}, the Vanish architecture
is vulnerable to Sybil attacks~\cite{Defeat-Vanish}.
Authors of Vanish have proposed countermeasures to this specific attack \cite{Vanish-Corrections}.
However, as discussed in~\cite{Defeat-Vanish}, this has only raised the bar for the attacker, due to the vulnerability of DHTs to Sybil attack,
thus, leaving as open question whether or not DHTs are the best choice for key-share storage.

Observe that the main flaw of Vanish design is that DHTs are assumed to be resistant to
crawling. On the contrary, monitoring all DNS resolvers is realistically infeasible.
Further, EphPub provides several improvements in terms of robustness and usability. First, 
it allows users to select the expiration time: by choosing domain names matching desired lifetime period, 
users assign their content a more precise expiration time (while in Vanish this is assigned
by the specific DHT implementation and due to network churn, unless using a so-called ``refreshing proxy''
that re-pushes key shares). 
Also, as we show next, given its simple nature, EphPub does not require users to install any additional software (e.g., DHT client). Thus,
it can be easily ported on any architecture and deployed even on mobile phones.
Indeed, researchers are already advocating efficient solutions to guarantee retroactive privacy, for instance, 
in {\em location sharing} applications over smartphones~\cite{freudigerprivate}.

\section{Conclusion \& Future Work}\label{conclusion}
This paper formalized the concept of Ephemeral Publishing (EphPub) and presented a novel efficient and robust solution to it,
based on the caching mechanism of the DNS. 
It  is transparent to existing applications and services and lets the sender to control data's lifetime
with relatively finer granularity than prior work~\cite{Vanish}.
Our EphPub technique uses a Reed-Solomon error correction code to minimize the error ratio. 
Our experimental analysis attests to EphPub's efficiency and shows that the protocol correctly handles ephemeral keys. 
Measurements conducted over several hundred thousands open DNS resolvers may be of
independent interest. We estimated the number of resolvers suitable for EphPub to 1.7 million.

We also proposed a command-line tool and a Firefox and a Thunderbird extension prototypes
that implement EphPub. Our software improves, by design, the usability of previous work~\cite{Vanish}.
Our implementations use a list of selected DNS cache resolvers that minimize error rates.
Nonetheless, EphPub users can build their own set of random resolvers. 

EphPub guarantees that published content cannot be decoded after its expiration  time, i.e., after that, an adversary
cannot retrieve the decryption key, hence, she cannot reconstruct the content.
Note that this guarantee holds even if the adversary controls (or compromises) the
sender and the receivers, and tries to recover the key or the plaintext from their memory. 
Indeed, we assume that senders (resp., receivers) never store plaintexts 
and encryption/decryption keys,  after publishing (resp., reading) the content, and before the expiration time.
While one may consider this to be a ``strong'' assumption, 
observe that EphPub's retroactive privacy guarantees are useful even without this assumption.
In fact, EphPub ensures that expired content cannot be retrieved by the types of attackers
(similar to {\em honest-but-curious} adversaries), envisioned in EphPub applications.
For instance, recall the examples from Section~\ref{Introduction}: even if 
plaintexts are not erased by the publisher, an employer is still prevented from crawling
social networks, blogs and forums, or cached data to access expired content published by his employees.

We emphasize that EphPub does not protect against a user that copies
or redistributes decrypted content, although the EphPub software is designed not to
store it on local memory.

EphPub represents an initial foray into DNS-based Ephemeral Publishing, thus, much remains to be done. 
Future work includes a deeper analysis of existing DNS monitoring systems,
a large-scale evaluation on the DNS load, and investigating alternative approaches based
on other caching mechanisms of today's Internet (e.g., web caching).

We also acknowledge that, if EphPub was to be used by a high number (e.g., millions) of users, 
it could impose a remarkable and unwanted traffic on the (open) DNS cache resolvers.
Thus, it is unclear how resolvers' administrators will react.
Nevertheless, we hope that the privacy protection offered by systems like EphPub will result
into a community effort, where users may contribute by providing DNS resources.
One prominent example of this practice is the Tor Project~\cite{TOR}, where a volunteer network
of servers route Internet traffic in order to conceal a user's location or web usage.

\descr{Acknowledgments.}
The authors would like to thank 
Daniele Perito for providing us the initial list of DNS servers, Saghar Estehghari for working on the extensions implementation, as well as James Griffioen, Vincent Jug\'e
and the ICNP 2011 anonymous reviewers for their insightful comments and help to
improve this article.


\end{document}